# Note concerning a high spectral resolution slicer for imaging spectro-polarimetry with the new generation Multichannel Subtractive Double Pass (MSDP) onboard the future EST telescope

*Malherbe, J.-M. (1), Sayède, F. (2), Mein P. (3), Roudier Th. (4)*

*Emérite LESIA/OP (1), GEPI/OP (2), Honoraire LESIA/OP (3), IRAP/OMP (4)*

*November 8, 2023*

## Abstract

Imaging spectroscopy is intended to be coupled with adaptive optics (AO) on large telescopes, such as the future European Solar Telescope (EST), in order to produce high spatial and temporal resolution measurements of velocities and magnetic fields upon a 2D field of view. We propose a high spectral resolution slicer (30 mÅ typical) for the Multichannel Subtractive Double Pass (MSDP) of EST, using a new generation slicer for thin photospheric lines such as those of FeI (56 channels, 0.13 mm step) which will benefit of AO and existing polarimeters. The aim is to reconstitute **cubes** of **instantaneous data (X, Y, lambda)** at high cadence, allowing the study of the photospheric dynamics and magnetic fields.

## Keywords

imaging spectroscopy, polarimetry, MSDP, solar physics, photosphere, chromosphere

## 1 – Introduction

The Multichannel Subtractive Double Pass is an imaging spectroscopy technique introduced by Mein (1977). It is based on a slicer which provides line profiles with N sampling points (or N channels) over a 2D field of view (FOV); for that purpose, the MSDP uses a rectangular entrance window instead of a thin slit. The technique was progressively developed and implemented on many telescopes (Mein *et al*, 2021). The first instrument with N = 7 channels was incorporated to the 14 m spectrograph of the Meudon Solar Tower. It was mainly running with the Hα line in order to study the dynamics (via the Doppler effect) of chromospheric features. It was followed by the MSDP of the Pic du Midi Turret Dome (Mein, 1980) with N = 11 channels and a better spectral resolution. The third instrument was integrated to the 15 m spectrograph of the german Vacuum Tower Telescope (VTT) in Tenerife (Mein, 1991). Meanwhile, polish colleagues introduced the technique on the large Bialkow coronagraph to observe prominences (Rompolt, 1994). The next instrument was put inside the 8 m spectrograph of the THEMIS telescope (Mein, 2002). For the first time, the MSDP worked there in polarimetric mode using achromatic waveplates and a calcite beam-splitter. Mein *et al* (2009) used the full Stokes capabilities of this high performance MSDP (N = 16) to investigate the vector magnetic field above an active region.

The slicers of these first generation instruments were based on multi-slits in the spectrum and prism beam-shifters. However, this technique does not allow to increase much the spectral resolution (limited to about 80 mÅ) and the number of channels. For that reason, the second generation of slicers uses now micro-mirrors, which allow to reach about 30 mÅ of spectral resolution and can deliver more than 50 channels. The first slicer using this technology was installed at the Meudon Solar Tower (N = 18). The Solar Emission Line Dopplerometer, in project with polish, british and slovakian colleagues for the Lomnicky coronagraph (N = 24) is dedicated to velocity measurements in the hot corona using the forbidden lines of iron (Malherbe *et al*, 2021) and is intended to operate in 2025. The future EST telescope could benefit of a state-of-the-art MSDP with N = 56 channels, as suggested by Sayède (2023) and Malherbe *et al* (2023). Two slicers were proposed for chromospheric and photospheric lines providing spectral resolutions in the range 40-100 mÅ. The present paper discusses the opportunity of an optional slicer with a special resolution of 30 mÅ (or less) for thin photospheric lines, such as those of FeI.

## 2 – Capabilities of the optional high spectral resolution slicer

The MSDP project for the future EST is described by Sayède (2023) and the optical capabilities are summarized by Malherbe *et al* (2023). It includes two 56-channel slicers inside the 8-meter spectrograph. The micro-mirrors step is Δx = 0.18 mm and 0.27 mm, corresponding to spectral resolutions (depending on the grating dispersion) in the range 40-100 mÅ, well adapted to photospheric and chromospheric lines.

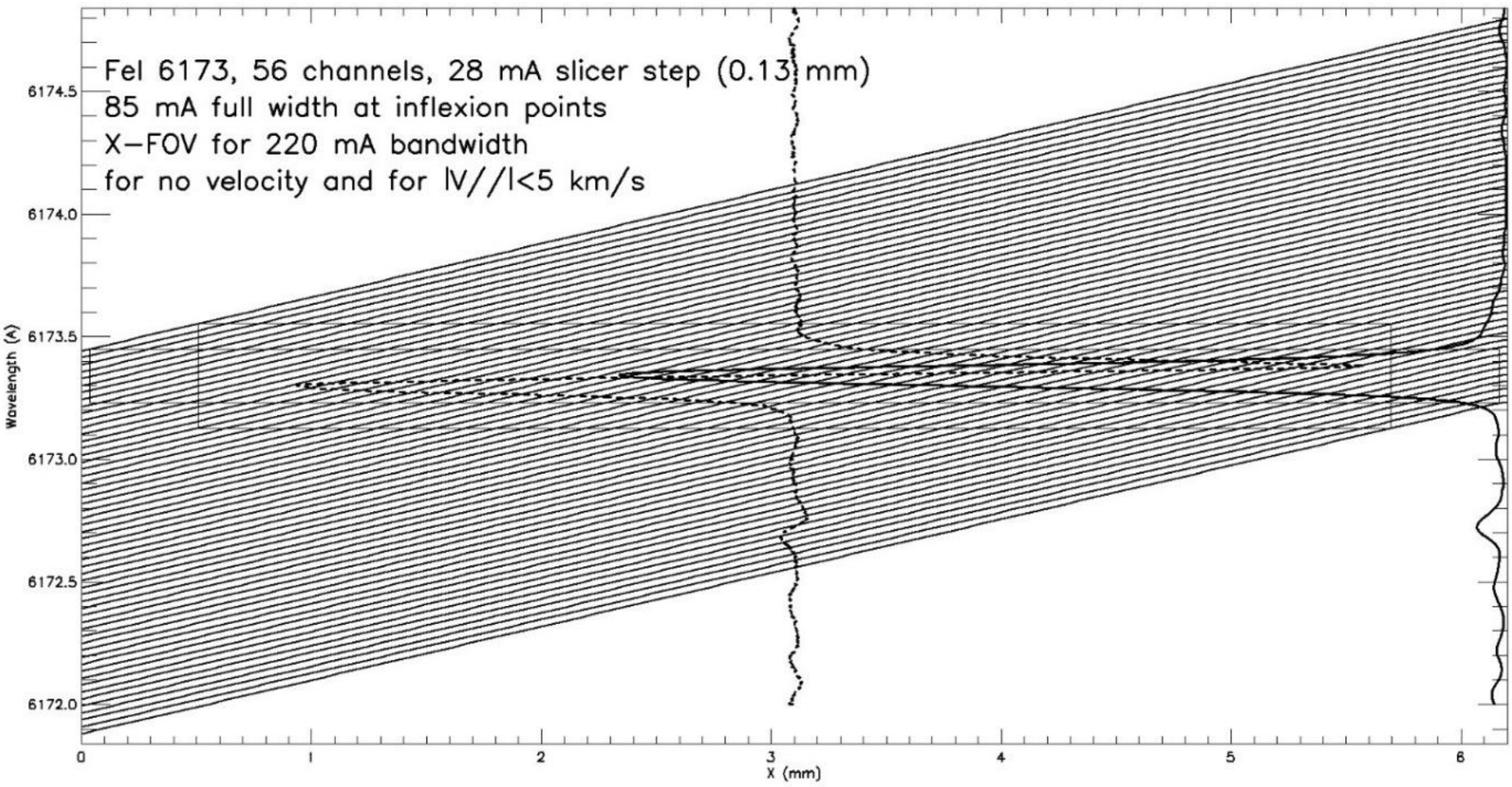

***Figure 1**: the MSDP wavelength transmission $\lambda_n(x)$ for the 56 channels (0.028 Å spectral resolution) for FeI 6173. The line profile at disk centre and its first derivative are displayed. For a given bandpass of 220 mÅ centred on the line core, the graph shows that the full FOV ($x_m$ = 6.2 mm = 8" in x-direction) is covered for no velocity and that 85% of the FOV is covered for velocities less than 5 km/s.*

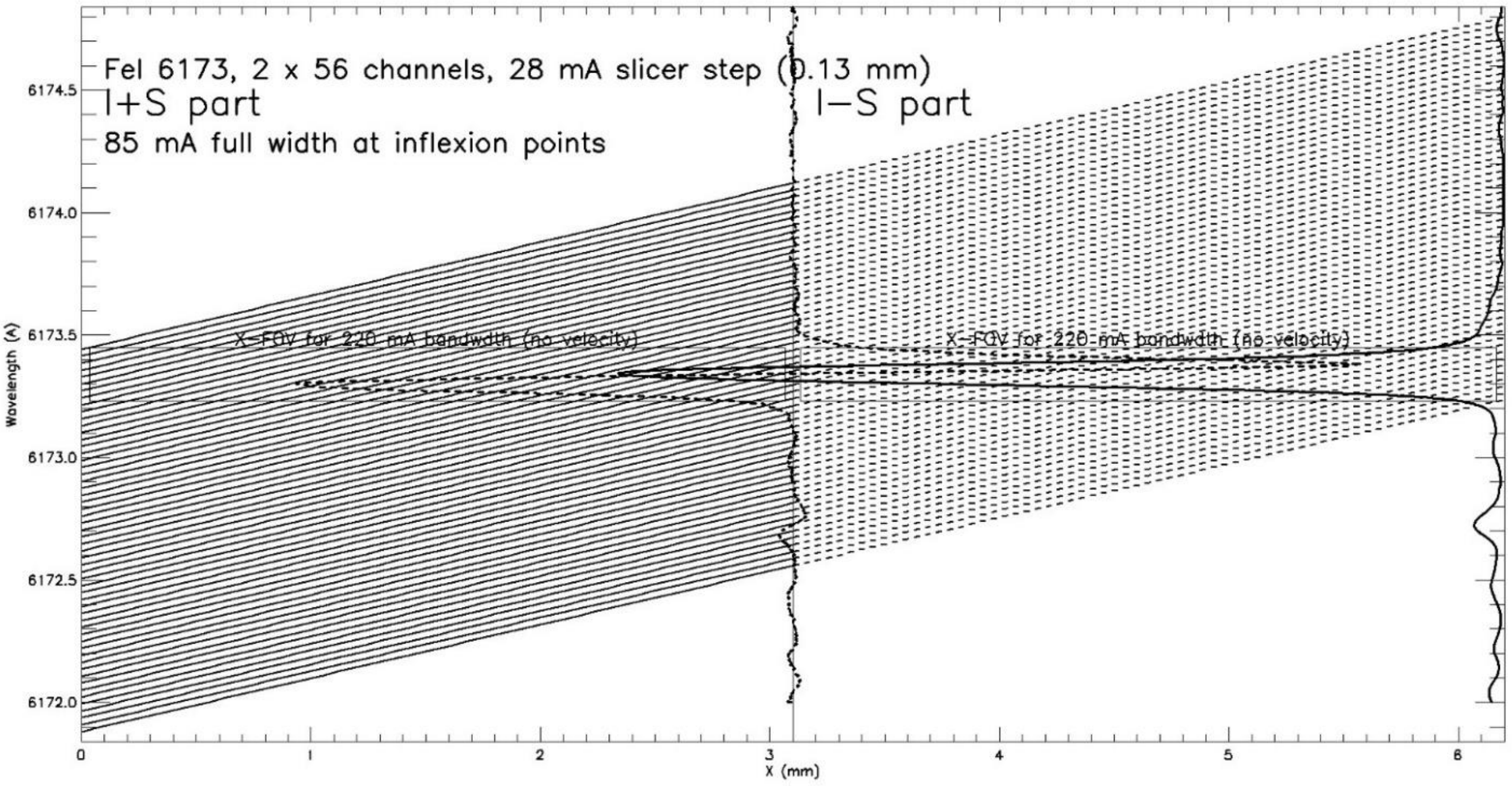

***Figure 2**: the MSDP wavelength transmission $\lambda_n(x)$ for the 56 channels (0.028 Å spectral resolution) for FeI 6173 and for simultaneous polarimetric measurements I+S and I-S (S = Q, U, V in sequence) with the method*

*described by Malherbe et al (2023). The FOV is two times smaller in x-direction (3.1 mm instead of 6.2 mm, or 4" instead of 8") in order to observe simultaneously I+S and I-S which are cospatial. The line profile at disk centre and its first derivative are displayed. The bandpass centred on the line core is 220 mÅ.*

However, for specific requirements concerning thin photospheric lines such as those of FeI, a slicer with a better spectral resolution could be proposed as an option. The step of such a slicer would be about $\Delta x = 0.13$ mm, providing spectral resolutions of 30 mÅ in the red part of the spectrum (28 mÅ for FeI 6173, Figures 1 & 2), and even better in the blue part.

Figure 1 shows the wavelength transmission of the 56 channels with no polarimetry, or in the case of polarimetry using a high speed modulator providing successive measurements of I-Q, I+Q, I-U, I+U, I-V, I+V (at high cadence over a stabilized FOV by the AO system). The sampling points are given by $\lambda_n(x) = \lambda_0 + (x/d) + n\,(\Delta x/d)$, for $0 < x < x_m$ ($x_m = 6.2$ mm is the maximum FOV in x-direction, it corresponds to 8"). In this formula, n is the current channel ($1 < n < N$, $N = 56$), d is the dispersion of the spectrograph (mm/Å), $\Delta x$ the slicer step (mm, the spectral resolution in Å is $\Delta x/d$). Generally, a cubic interpolation between sampling points is used for the determination of precise line shifts produced by the Doppler or Zeeman effects.

On the contrary, Figure 2 displays the wavelength transmission of the channels with polarimetry using a dual beam polarimeter (simultaneous measurements of I-S and I+S, with S = Q, U, V sequential). In that case, the FOV is divided by 2 by a mask which allows to form I-S at left ($0 < x < x_m/2$) and I+S at right ($x_m/2 < x < x_m$). Please notice that both regions are strictly cospatial on the Sun, but for a given abscissa in x-direction, such that $0 < x < x_m/2$, the line profiles of I-S and I+S use different sampling points, respectively given by $\lambda_n(x) = \lambda_0 + (x/d) + n\,(\Delta x/d)$ and $\lambda_n(x) = \lambda_0 + [(x + x_m/2)/d] + n\,(\Delta x/d)$. For that reason, interpolations are necessary to restore the same sampling points, in order to compare I-S and I+S profiles for Dopplershift and magnetic field calculations.

## 3 – Simulation of the channels provided by the high resolution slicer in the case of full FOV sequential measurements of I+S (S = -Q, +Q, -U, +U, -V, +V) using the FeI 6173 spectral line

This section presents a simulation of the 56 channels of the MSDP in the case of the FeI 6173 line. It is the line used by the HMI instrument onboard the SDO satellite, and also the line used by the PHI instrument onboard the SOLAR ORBITER mission. This is a thin line (85 mÅ for the Full Width at Inflexion Points, FWIP) very sensitive to the photospheric magnetic field (equivalent Landé factor 2.5). The depth of the line lies around 63 % of the continuum in the quiet Sun. SDO produces, in this line, high cadence magnetograms of the full Sun (4000 x 4000 pixels) every minute (pixel size = 0.5") with moderate spectral resolution (6 sampling points), assuming that the line profile has a gaussian shape (which helps a lot to derive Dopplershifts and magnetic fields).

The MSDP onboard the European Solar Telescope (EST) could work using, at least, two polarimetric modes already discussed by Malherbe *et al* (2023) and detailed by Figure 3 :

- High cadence modulation (100 Hz), allowing to observe in sequence I+Q, I-Q, I+U, I-U, I+V, I-V with adaptive optics providing image stabilization ; in that case, the full FOV of the MSDP is preserved (8" in x-direction, 120" in y-direction).
- Simultaneous observations of I+S and I-S (S = Q, U, V in sequence) of the half FOV (4" in x-direction, 120" in y-direction) using a mask (grid) and a birefringent beam-splitter. This method has the advantage of simultaneous polarimetric measurements, but the disadvantage of FOV reduction, so that two successive observations are required to record the full FOV.

Please notice that the full FOV at the telescope focus is a square of 32" x 30" ; for the MSDP, it is optically split in 4 parts which are rearranged to form a rectangle of 8" x 120", as shown by Figure 3 (left), before injection into the spectrograph. For simultaneous measurements of I+S and I-S signals with a calcite beam splitter, a mask is used (Figure 3, right) under the form of a grid of 8" step (method suggested by Semel, 1980). The mask covers the half FOV, so that two consecutive observations are needed to record the full FOV.

The following figures simulate the two polarimetric methods using SDO/HMI data of 15 September 2023 at 12:00 UT, such as the continuum intensity, dopplergram and Line Of Sight (LOS) magnetogram.

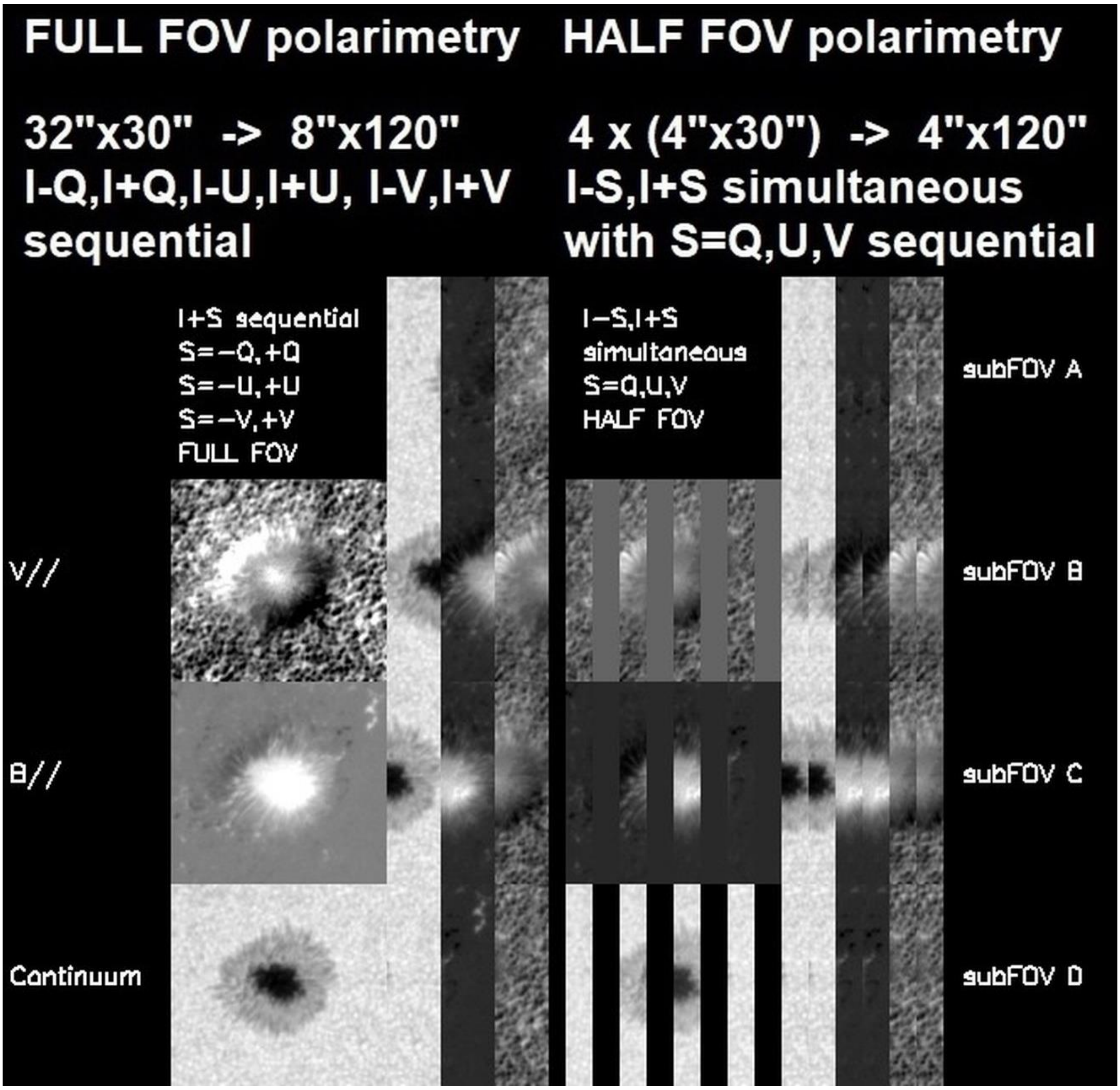

***Figure 3**: simulation of the MSDP in polarimetric mode using SDO/HMI data (continuum intensity, LOS magnetic field B// and LOS velocity field V//) of 15 September 2023, 12:00 UT. We start with a 32" x 30" FOV at the F/50 focus of the EST telescope. This FOV is transformed into a 8" x 120" FOV for the MSDP by splitting the primary image in 4 parts (A, B, C, D) which are optically re-aligned vertically before injection into the spectrograph. The fast modulation method (left) delivers sequential measures of I+S (S = -Q, +Q, -U, +U, -V, +V) over the full FOV. The beam-splitting method (right) requires grid masking, delivering the half FOV, but in that case I+S and I-S (S = Q, U, V in sequence) are simultaneous. A second observation is needed to observe the parts masked by the grid during the first observation.*

Figures 4 & 5 show a simulation of MSDP observations with the Δx = 0.13 mm/N = 56-channel slicer, that could be got using the fast modulation method, preserving the full FOV, but with I+V and I-V not simultaneous

(but very close in time). Figure 4 displays I+V (image_1), while Figure 5 shows I-V (image_2), which must be observed at high cadence (100 Hz) with adaptive optics turned on. The wavelength sampling points are those of Figure 1. The circular polarization rate V/I = (Image_1 – Image_2) / (Image_1 + Image_2) is shown by Figure 6, from which it is easy to deduce the LOS magnetic field B// using the weak field theory. An alternative method consists in measuring the wavelength shift between the I+V and the I-V line profiles, as this shift is proportional to the LOS magnetic field B//. Both I+V and I-V profiles are globally translated of the same quantity in case of Dopplershifts V// of the plasma.

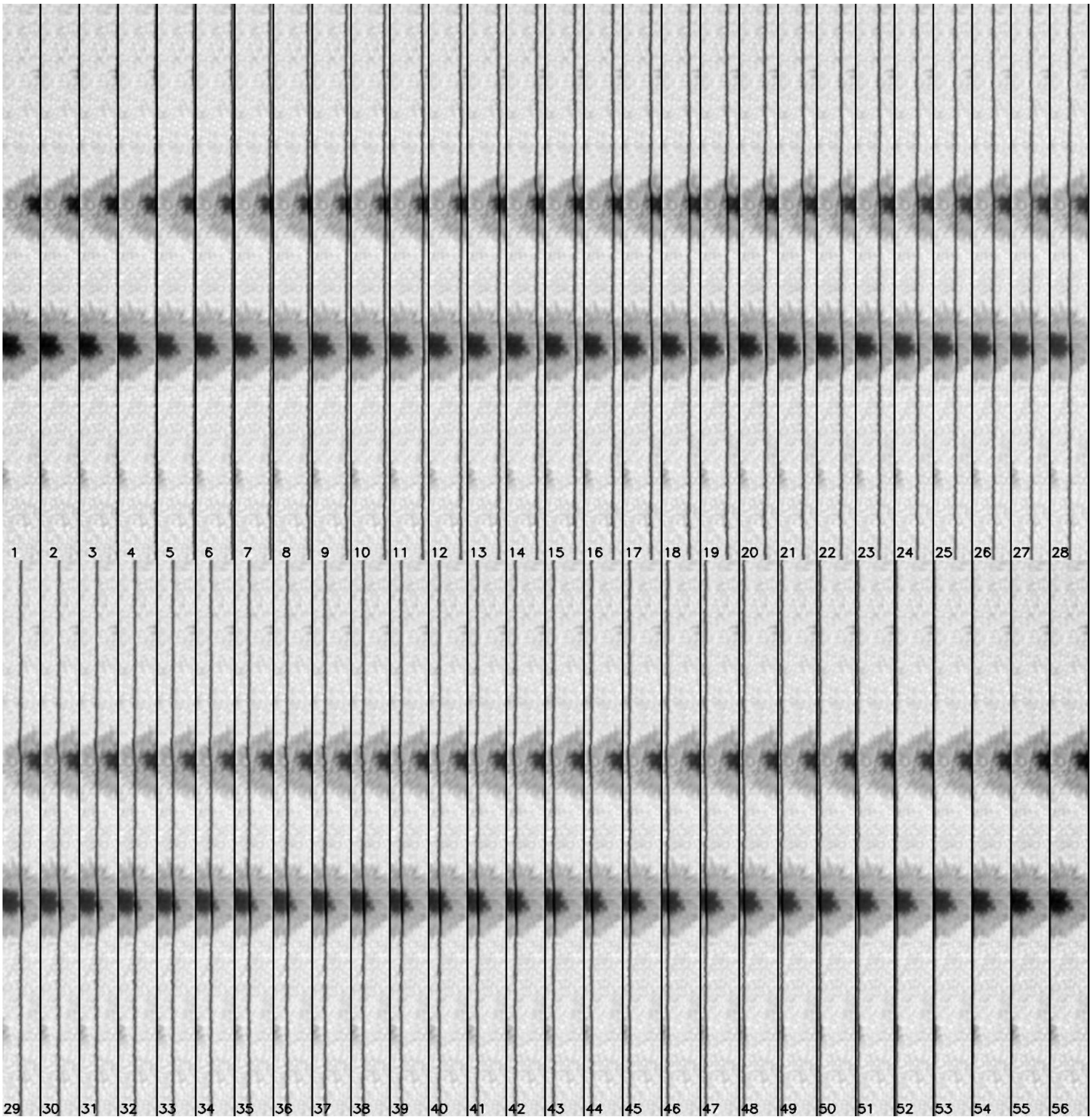

*Figure 4: simulation using SDO data (continuum, LOS magnetic field B//, LOS velocity field V//) of the 56-channel slicer (28 mÅ resolution, Δx = 0.13 mm) with FeI 6173 line (assumed gaussian), here the I+V spectra image (image_1). I+V profiles can be reconstructed using the chart of Figure 1. The FOV is 8" x 120".*

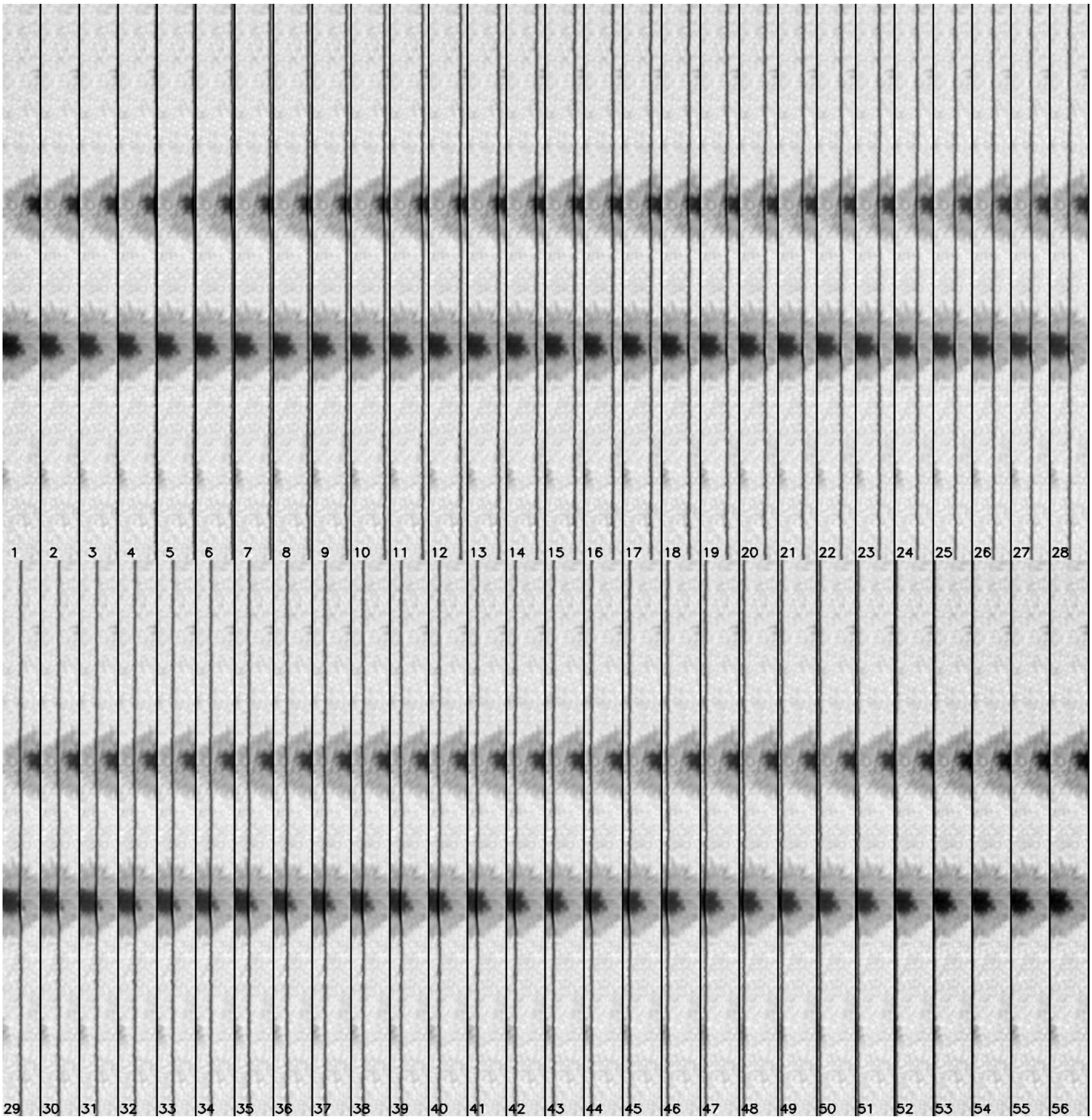

***Figure 5**: simulation using SDO data (continuum, LOS magnetic field B//, LOS velocity field V//) of the 56-channel slicer (28 mÅ resolution, Δx = 0.13 mm) with FeI 6173 line (assumed gaussian), here the I-V spectra image (image_2). I-V profiles can be reconstructed using the chart of Figure 1. The FOV is 8" x 120", it corresponds to the telescope FOV of 32" x 30" after MSDP re-arrangement.*

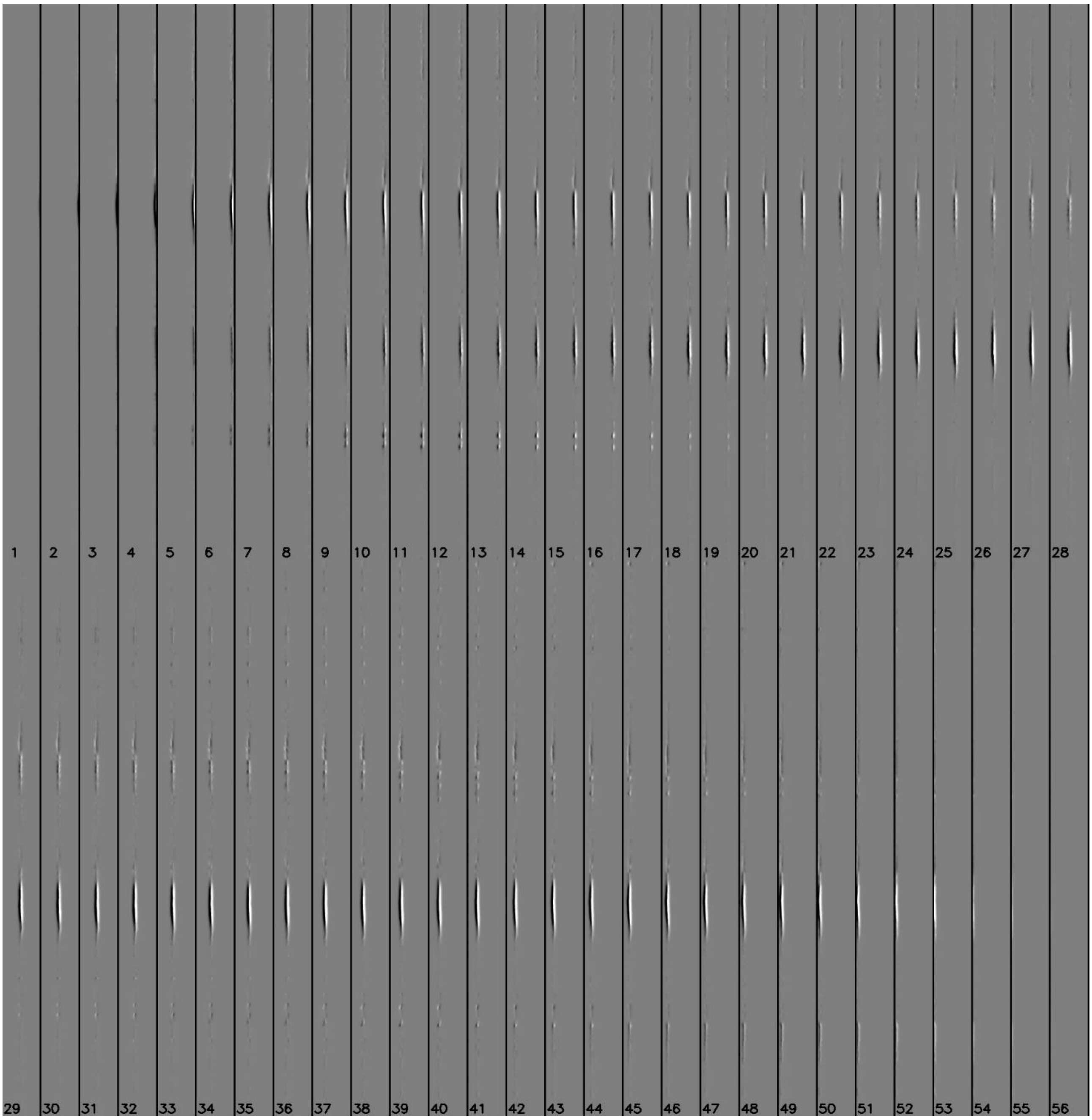

***Figure 6***: *simulation using SDO data (continuum, LOS magnetic field B//, LOS velocity field V//) of the 56-channel slicer (28 mÅ resolution, Δx = 0.13 mm micro-mirrors) with FeI 6173 line (assumed gaussian). Here, we display the circular polarization rate V/I = (image_1 - image_2) / (image_1 + image_2), where image_1 (I+V) and image_2 (I-V) are those of figures 4 & 5. I+V and I-V line profiles can be reconstructed using the wavelength chart of Figure 1 and have exactly the same sampling points. The weak field theory was used for the simulation.*

Figure 7 shows the simulation of MSDP line profiles, in polarimetric mode over the full FOV derived from sequential observations of I+V, I-V, that could be delivered by the 56-channel slicer with a high cadence modulator. The sampling points are those given by Figure 1 for 3 locations in the x-direction of the FOV, x = 0, x = $x_m$/2 and x = $x_m$ ($x_m$ = 6.2 mm or 8”). The profiles of the circular polarization rate V/I are also displayed. Figure 8 shows the simulation of line profiles derived from sequential observations of I+Q, I-Q, that could be got with the same modulator. The sampling points are also those of Figure 1 for the 3 locations in the x-direction. The profiles of the linear polarization rate Q/I are indicated.

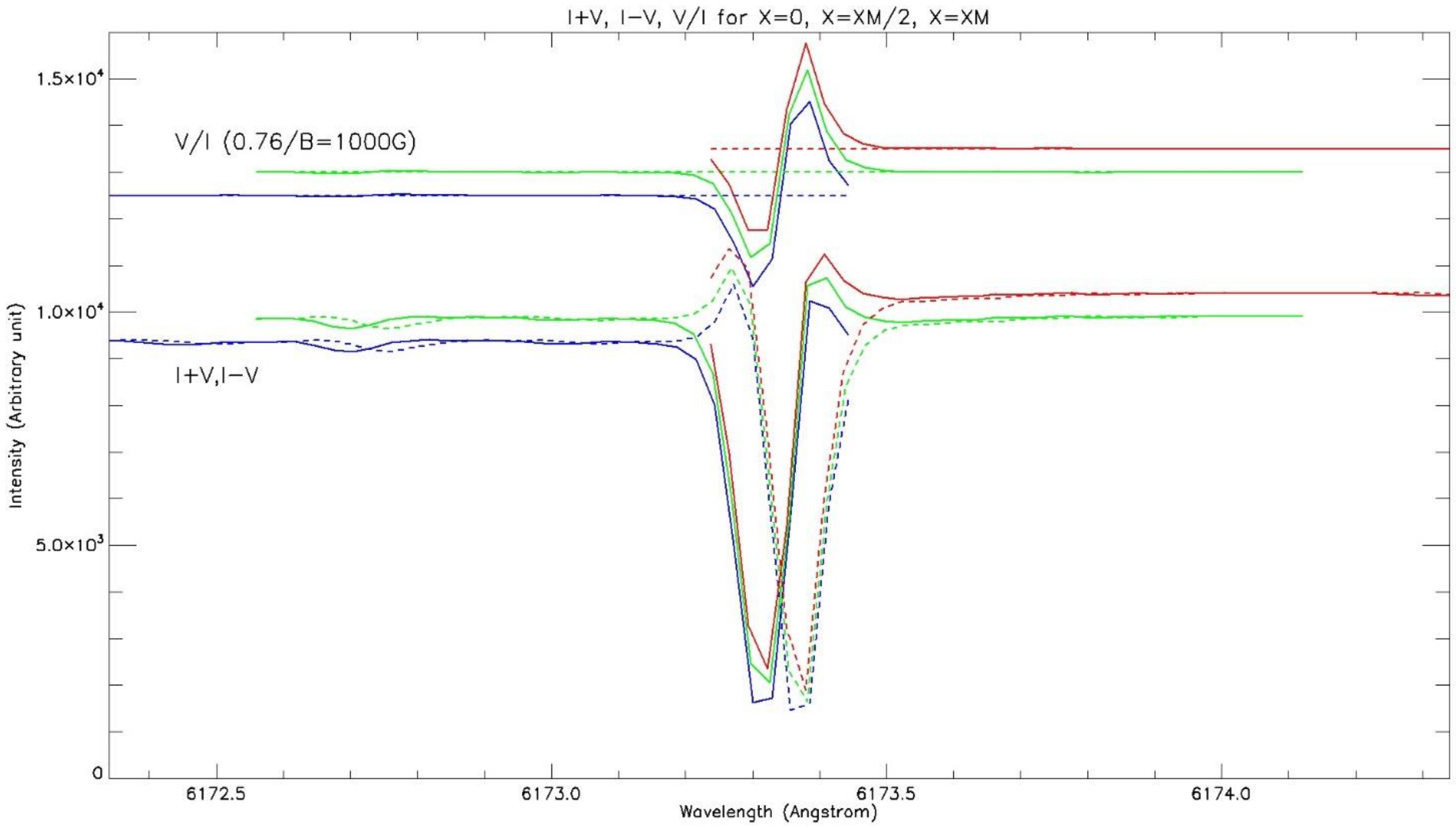

*Figure 7: simulation of I+V (solid line) and I-V (dashed) line profiles of FeI 6173 (bottom), that could be delivered by the Δx = 0.13 mm / N= 56-channel slicer. The spectral resolution is 28 mÅ. Line profiles were computed for $B_{//}$ = 1000 G. Profiles for 3 locations in the FOV are drawn, x = 0 (left FOV, blue), x=$x_m$/2 (centre, green) and x = $x_m$ (right, red) for $x_m$ = 8" (or 6.2 mm). Sampling points are raw data (no interpolation between them). Top: the V/I circular polarization rate (theoretical maximum 0.76 for 1000 G).*

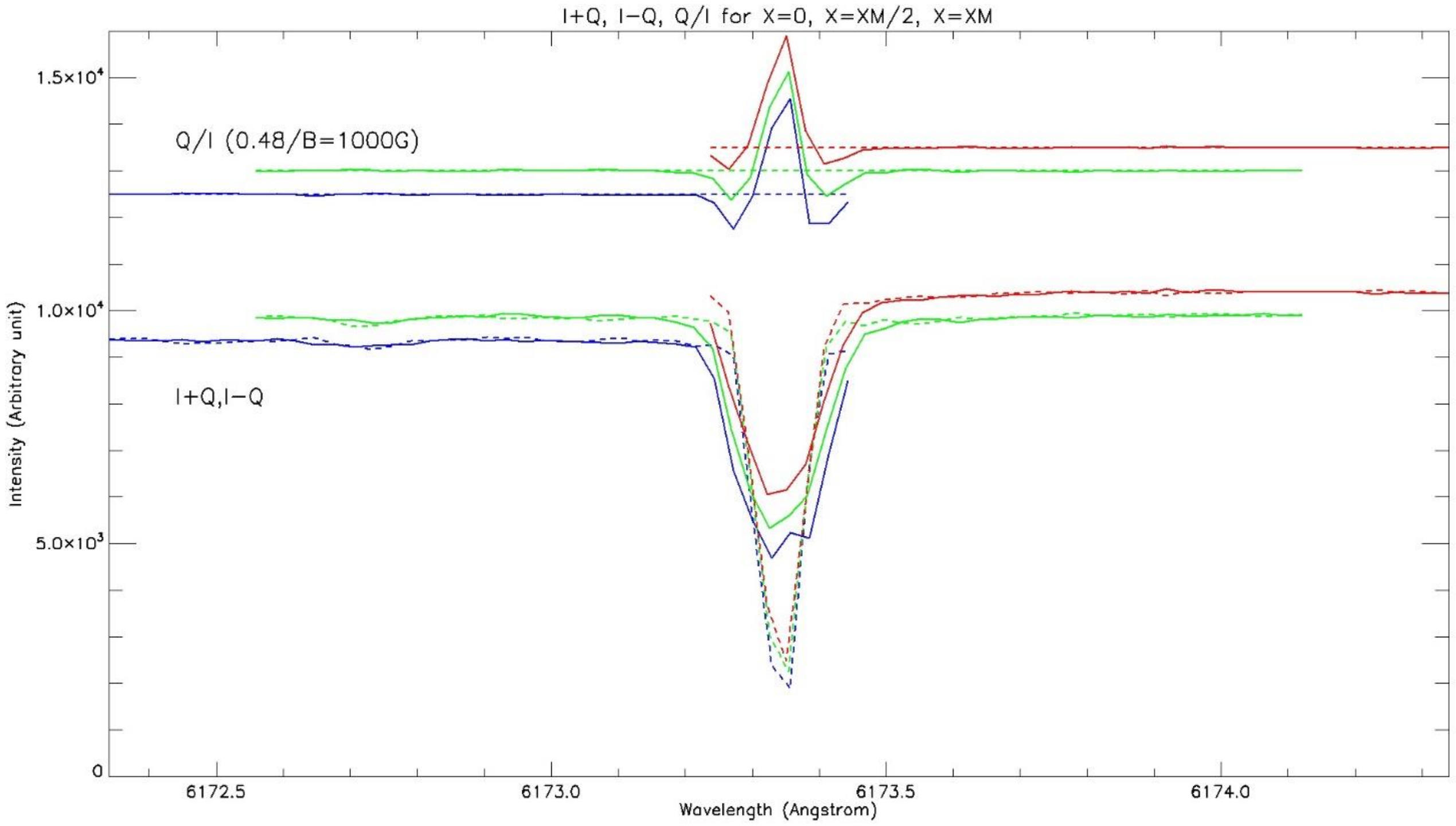

*Figure 8: simulation of I+Q (solid line) and I-Q (dashed) line profiles of FeI 6173 (bottom), that could be got using the Δx = 0.13 mm / N= 56-channel slicer. The spectral resolution is 28 mÅ. Line profiles were computed for $B_\perp$ = 1000 G. Profiles for 3 locations in the FOV are drawn, x = 0 (left FOV, blue), x=$x_m$/2 (FOV centre, green) and x = $x_m$ (right FOV, red) for $x_m$ = 8" (or 6.2 mm). Sampling points are raw data (no interpolation between them). Top: the Q/I linear polarization rate (theoretical maximum 0.48 for 1000 G).*

### 4 – Simulation of the channels provided by the high resolution slicer in the case of half FOV simultaneous measurements of I+S and I-S (S = Q, U, V sequential) using the FeI 6173 spectral line

Figure 9 shows a simulation of MSDP observations with the 56-channel slicer, that could be obtained using simultaneous observations of I+V and I-V of the half FOV with a birefringent beam-splitter and a mask (grid). Each channel is divided into two subchannels showing the same half FOV (4" x 120"). I+V and I-V are strictly cospatial and simultaneous, and their profiles can be reconstructed using the chart of Figure 2. In order to recover the full FOV of 8" x 120", corresponding to the square FOV of 32" x 30" split in 4 parts at the focus of the telescope, a second observation is needed and requires to move the telescope of 4" in the x-direction.

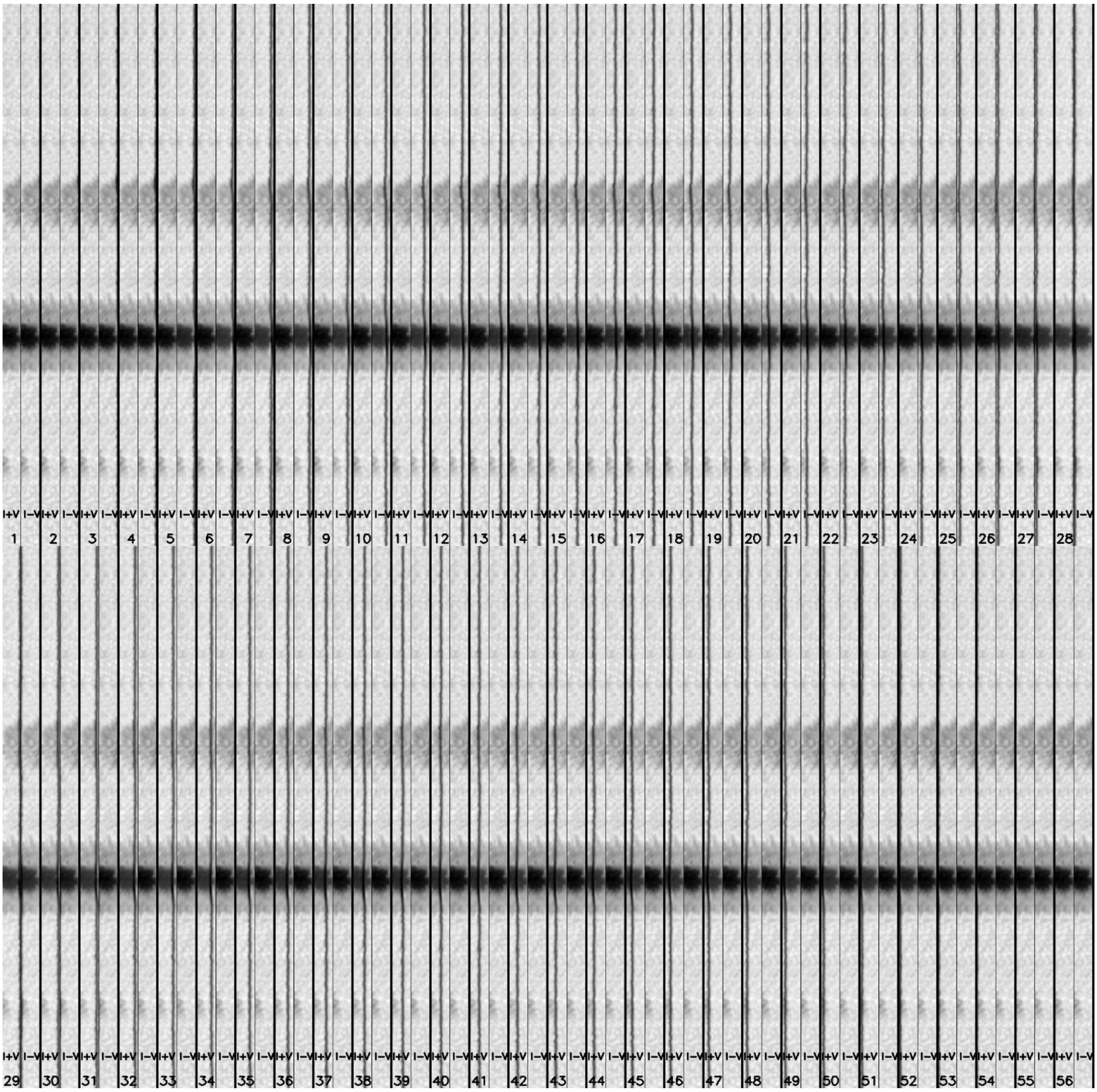

***Figure 9**: simulation using SDO data (continuum, LOS magnetic field B//, LOS velocity field V//) of the 56-channel slicer (28 mÅ resolution or 0.13 mm) with FeI 6173 line (assumed gaussian). The calcite beam splitter and the grid produce 56 x 2 = 112 cospatial subchannels. I+V and I-V are simultaneous, implying the FOV reduction by a factor 2 in the x-direction (4" x 120" instead of 8" x 120"). I+V and I-V line profiles can be reconstructed using the chart of Figure 2 for the two cospatial subchannels.*

Figure 10 shows the simulation of MSDP line profiles, in polarimetric mode over the half FOV with a birefringent beam-splitter, derived from simultaneous observations of the cospatial subchannels I+V, I-V, that could be obtained using the 56-channel slicer. The wavelength sampling points are those provided by Figure 2 for 3 locations in the x-direction of the half FOV, x = 0, x = $x_m$/4 and x = $x_m$/2 ($x_m$/2 = 3.1 mm or 4”). The profiles of the circular polarization rate V/I are also displayed. Figure 11 shows the simulation of line profiles derived from simultaneous observations of I+Q, I-Q, in the same configuration. The sampling points are also those of Figure 2 for the 3 locations. The profiles of the linear polarization rate Q/I are indicated.

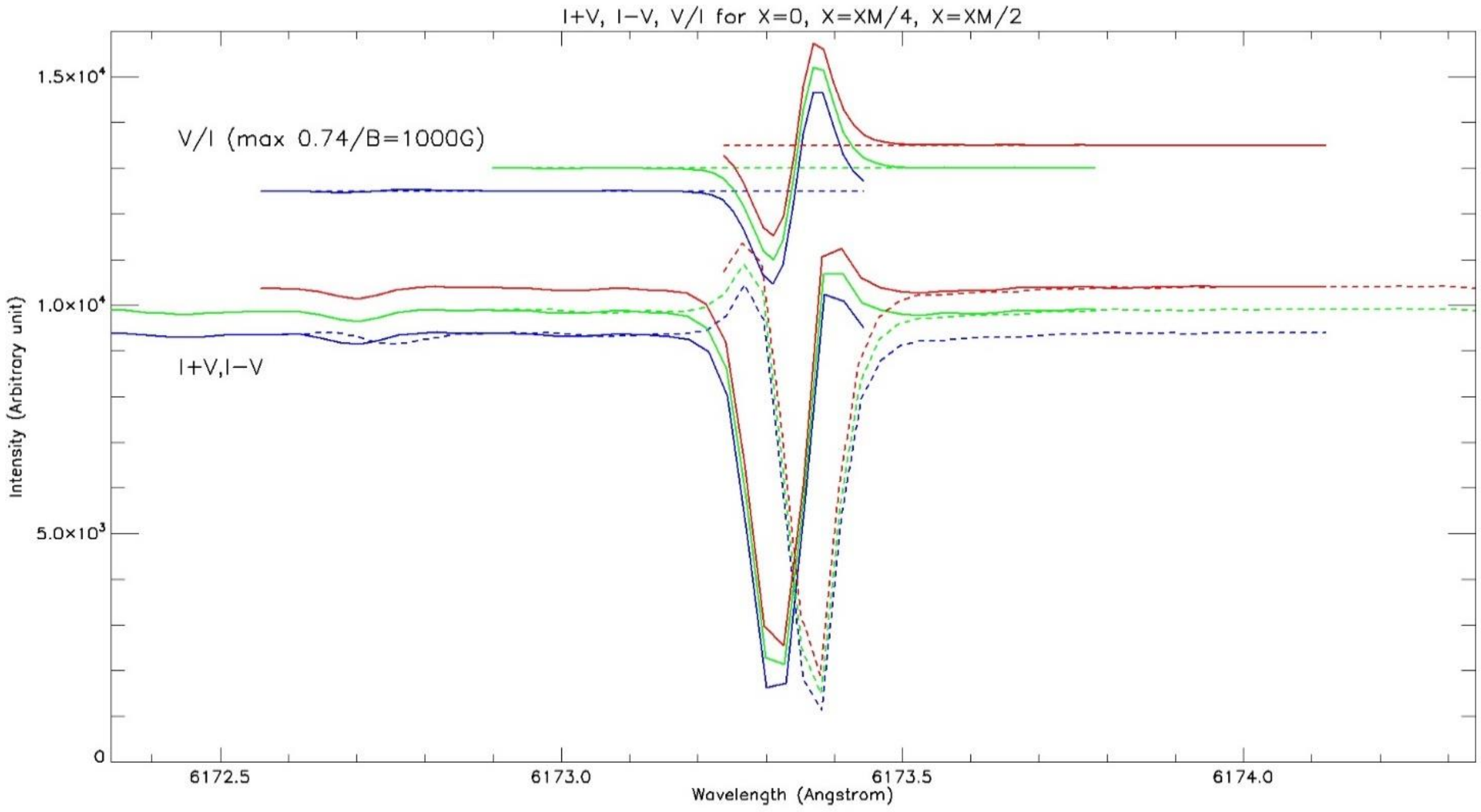

***Figure 10**: simulation of I+V (solid line) and I-V (dashed) line profiles of FeI 6173 (bottom), that could be got using the Δx = 0.13 mm / N= 56-channel slicer. Profiles for 3 locations in the FOV are drawn, x = 0 (left FOV, blue), x=$x_m$/4 (centre, green) and x = $x_m$/2 (right, red) for $x_m$/2 = 4” (or 3.1 mm). Top: V/I polarization rate.*

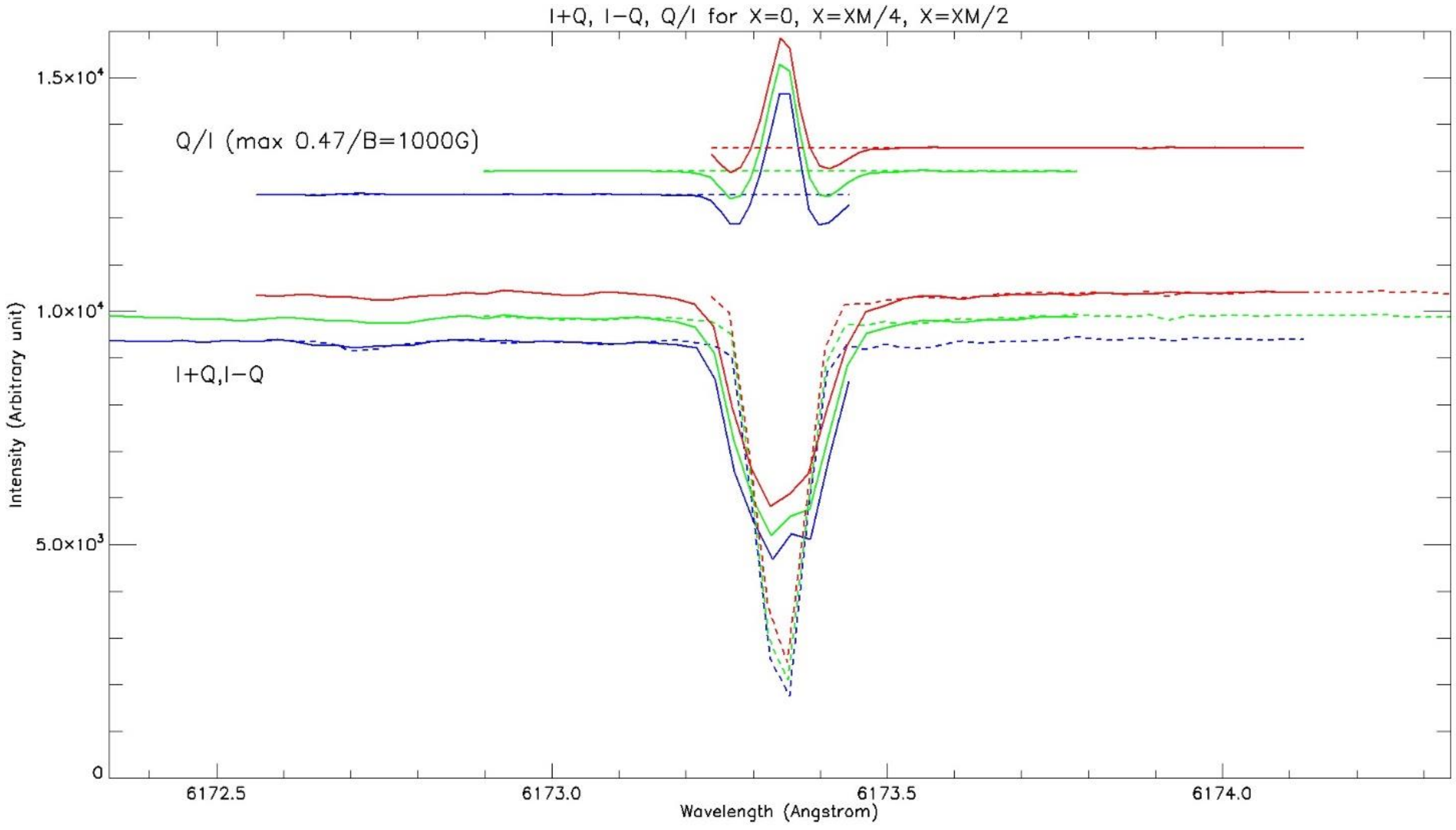

***Figure 11**: simulation of I+Q (solid) and I-Q (dashed) line profiles of FeI 6173 (bottom), that could be got using the Δx = 0.13 mm / N= 56-channel slicer. Profiles for 3 locations in the FOV are drawn, x = 0 (left FOV, blue), x=$x_m$/4 (centre, green) and x = $x_m$/2 (right, red) for $x_m$/2 = 4” (or 3.1 mm). Top: Q/I polarization rate.*

## 5 – Line shape discussion and wavelength sampling

This short section discusses the gaussian assumption that was used in the simulations. Figure 12 shows a gaussian fitting of the atlas profile at disk centre of FeI 6173 for different samplings (crosses), respectively 30, 75, 105 and 135 mÅ. We found that the gaussian fitting of this almost symmetric profile works well for sampling better than 60 mÅ, so that a high resolution slicer (0.13 mm) is not necessary in that special and restrictive case, for which the standard slicers (0.18 mm and 0.27 mm steps) are convenient. Indeed, the fitting of lines close to the gaussian shape does not require very high spectral resolution, as demonstrated by the high quality of SDO/HMI data.

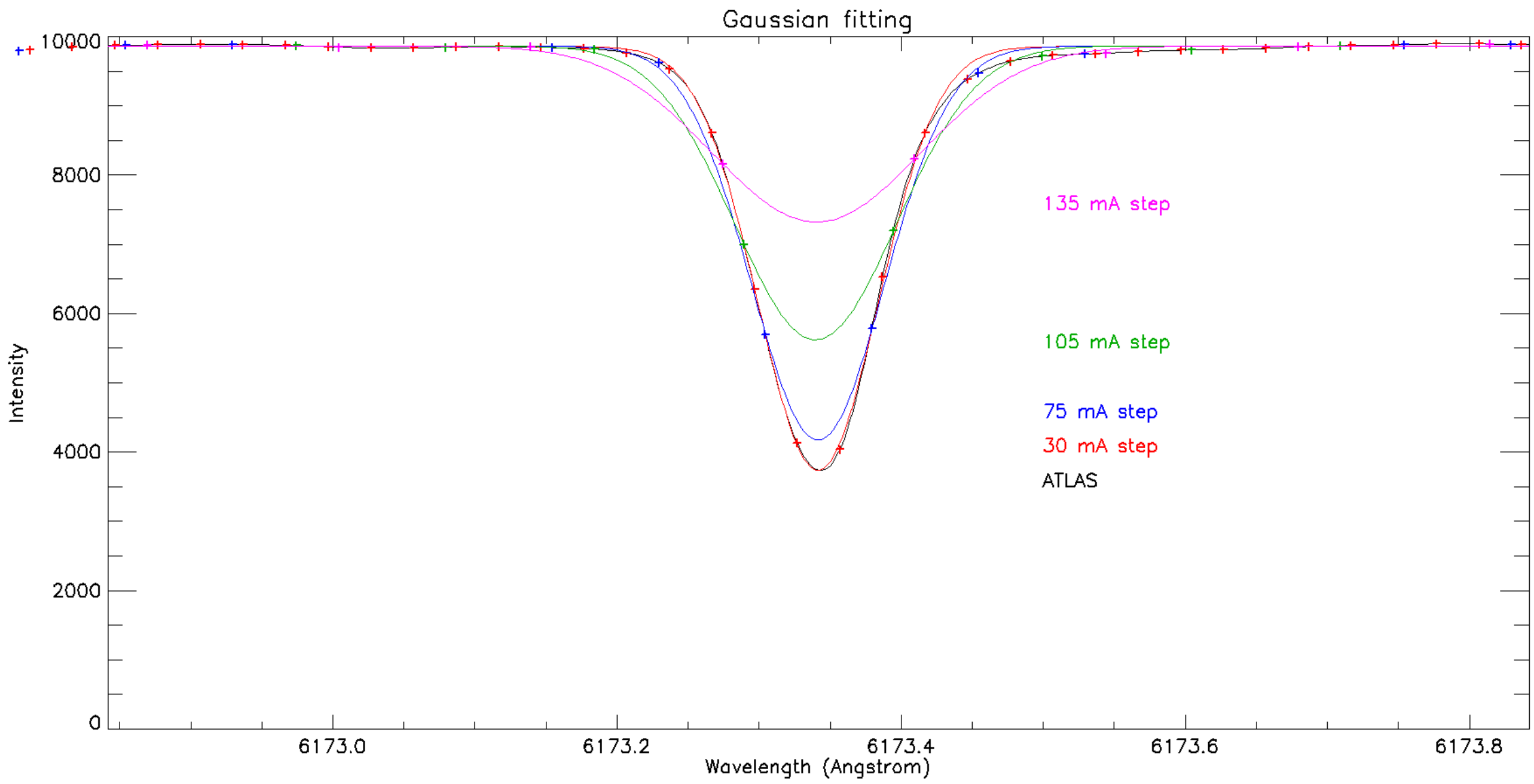

*Figure 12: gaussian fitting of the atlas profile (black) of FeI 6173 at disk centre for various samplings (30, 75, 105 and 135 mÅ), for respectively the red, blue, green and violet curves.*

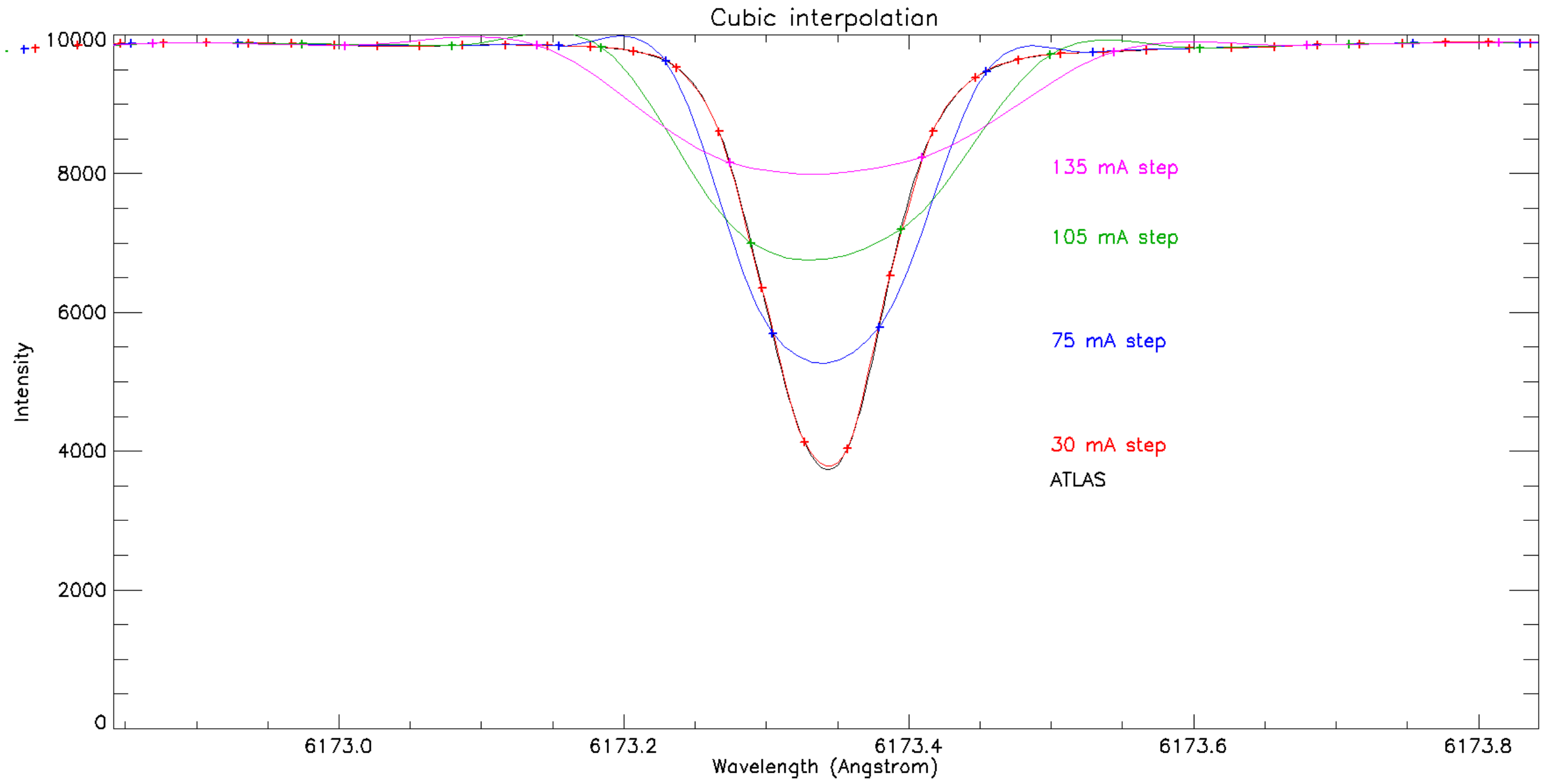

*Figure 13: cubic interpolation of the atlas profile (black) of FeI 6173 at disk centre for various samplings (30, 75, 105 and 135 mÅ), for respectively the red, blue, green and violet curves.*

Figure 13, on the contrary, does not make any hypothesis on the line shape, which can be asymmetric (this would be the case in a complex atmosphere) and far from a simple mathematical curve. A cubic interpolation was just done between the sampling points (crosses). In such a case, a high resolution slicer (0.13 mm) becomes necessary. The 30 mÅ sampling of the slicer provides excellent results, but above this value, the reconstruction of the profile is more and more degraded. For that reason, a high resolution slicer is a good option for thin and disturbed photospheric lines, as many of them (such as FeI 5250, 6173, 6302, CaI 6103) are very magnetically sensitive.

## 6 - Conclusion

The error at line centre, when the line has a nearly gaussian shape (this is the case of the atlas profile of FeI 6173 in the quiet Sun), remains below 1 % up to 60 mÅ sampling. When the line profile is asymmetric or far from a mathematical curve, the cubic interpolation is required to derive LOS velocities and magnetic fields ; it provides very good results (error less than 1 %) for sampling better than 30 mÅ. The SOT spectro-polarimeter onboard the HINODE satellite offered a similar resolution on FeI 6301 and 6302. Hence, a high resolution slicer for the MSDP onboard EST could be useful in order to avoid assumptions on the line shapes.